\documentclass[%
 reprint,
superscriptaddress,
 amsmath,amssymb,
prb,
floatfix,
]{revtex4-2}

\usepackage{xcolor}
\usepackage{graphicx}
\usepackage{dcolumn}
\usepackage{bm}
\usepackage{hyperref}

\DeclareMathOperator{\arccosh}{arccosh}

\begin{document}


\title{Laser beam focusing by layered superconductor \\ tuned by DC magnetic field}

\author{N. Kvitka}
    \email{kvitkanina@gmail.com}
\affiliation{O.~Ya.~Usikov Institute for Radiophysics and Electronics NASU, 61085 Kharkiv, Ukraine}

\author{H.V. Ovcharenko}%
\affiliation{V.~N.~Karazin Kharkiv National University, 61077 Kharkiv, Ukraine}

\author{Z.A. Maizelis}%
\affiliation{O.~Ya.~Usikov Institute for Radiophysics and Electronics NASU, 61085 Kharkiv, Ukraine}
\affiliation{V.~N.~Karazin Kharkiv National University, 61077 Kharkiv, Ukraine}

\author{S.S. Apostolov}%
\affiliation{O.~Ya.~Usikov Institute for Radiophysics and Electronics NASU, 61085 Kharkiv, Ukraine}

\author{V.A. Yampol’skii}%
\affiliation{O.~Ya.~Usikov Institute for Radiophysics and Electronics NASU, 61085 Kharkiv, Ukraine}
\affiliation{V.~N.~Karazin Kharkiv National University, 61077 Kharkiv, Ukraine}

\date{\today, v. 4}

\begin{abstract}
We develop a theory of the propagation and focusing of the THz Gaussian laser beam through the layered superconductor slab of finite thickness in the presence of an external DC magnetic field in a nonlinear regime. We show that, in this case, focusing of radiation, which results from the specific nonlinearity of the medium, can be flexibly tuned by the external magnetic field providing a new way to control THz waves. We analytically study the main characteristics of the Gaussian beam, its waist, and the focusing distance as the functions of wave frequency, amplitude, and the external magnetic field. The results of analytic calculations are supported by the numerical simulation of the electromagnetic field distribution in the slab.
\end{abstract}
\maketitle
\section{\label{sec:level1}INTRODUCTION}

THz radiation has been intensively studied during the last decades having a lot of applications in communication technologies, imaging, and medicine \cite{Gong2019, Saln2019, Pawar2013}. While THz emission mechanisms are well developed, focusing and control of radiation in this range still meets problems, including high role of diffraction and thus incomplete focusing, distortion of field distribution in the THz beam cross-section due to aberrations, high losses on the interfaces \cite{Pawar2013, Monnai:13, Rosanov2017}. Among materials used for conventional lens fabrication, Teflon and silicon are usually preferred due to their higher transmittance \cite{Wang2020, Hebling2008, Hafez2016}. Meanwhile, a lot of alternative ways of focusing THz waves were devised including creation of programmable stacks of waveguiding metal curved plates \cite{ Mendis2016}, cantilevered mechanical grating arrays \cite{ Monnai:13}, phased array antennas \cite{ Maki:08}, layered nonlinear materials \cite{ Weiss:01}. An important direction bases on using semiconductors which can be optically gated or on employing the lateral photo-Dember effect \cite{ Busch:12, Carthy2017}. The control of phase distribution in the beam can be achieved both optically and electronically \cite{ Carthy2017, Mendis2016}. Implementation of the layered superconductors to operate the THz radiation is yet another promising direction allowing flexible control and focusing of THz waves~\cite{ Savelev2010}.

Layered superconductors are periodic structures, in which superconducting layers are separated by the insulating ones. The examples include natural layered superconductors, such as $\rm Bi_2 Sr_2 Ca Cu_2 O_{8+\delta}$, and artificial structures, for example, $\rm Nb/Al-AlO_x/Nb$~\cite{ Savelev2010}. They have garnered significant attention for their unique physical properties explained by their strong anisotropic nature: electric transport along layers in such materials is similar to bulk superconductors, while in the perpendicular direction, it is determined by the Josephson effect of Couper pairs tunneling across the insulating barriers \cite{PhysRevB.93.075152, Hu2010}. This qualitative and quantitative anisotropy leads to the decrease of effective plasma frequency, the so-called Josephson plasma frequency, down to the THz range allowing to use these materials in THz technologies. Meanwhile, specific non-linear relation between the tunneling current and the phase parameter in the superconducting layers provides a wide variety of interesting phenomena such as self-induced transparency \cite{ PhysRevB.82.144521}, hysteretic response \cite{ PhysRevB.78.184504}, or specific superposition rules \cite{PhysRevB.90.184503}. An important property of layered superconductors is their high sensitivity to even weak DC magnetic field, at which no vortexes are formed, predicted for plane waves in Ref. \cite{hmain}. This nonlinear influence of static magnetic field on THz wave propagation can be used to control the wave transmittance.

Recently layered superconducting slabs were proposed to focus the THz Gaussian laser beams \cite{lasermain}. It was shown that the deviation from initial Gaussian profile of the field distribution in the beam cross-section can be neglected in some range of parameters, while the focusing ability can be very high and depends not only on the frequency of the radiation but also on its amplitude which can be used in applications. In this paper, we further develop this idea and show that the external DC magnetic field can be used to accurately tune the parameters of the beam even for the fixed frequency and amplitude.

The paper is organized as follows. In the second section, we describe our theoretical model and setup. In the third section, we study the effect of the external DC field on the main optical characteristics of laser beam focusing: the beam waist (its minimal radius) and the focal length. The subsection A of this section is devoted to the analytical approach, and in subsection B we present numerical simulation results which we compare with those obtained analytically. In the fourth section, we summarize the obtained results.

\section{\label{sec:level2}MODEL}

\begin{figure}[h]
\begin{center}
\includegraphics[width=7 cm]{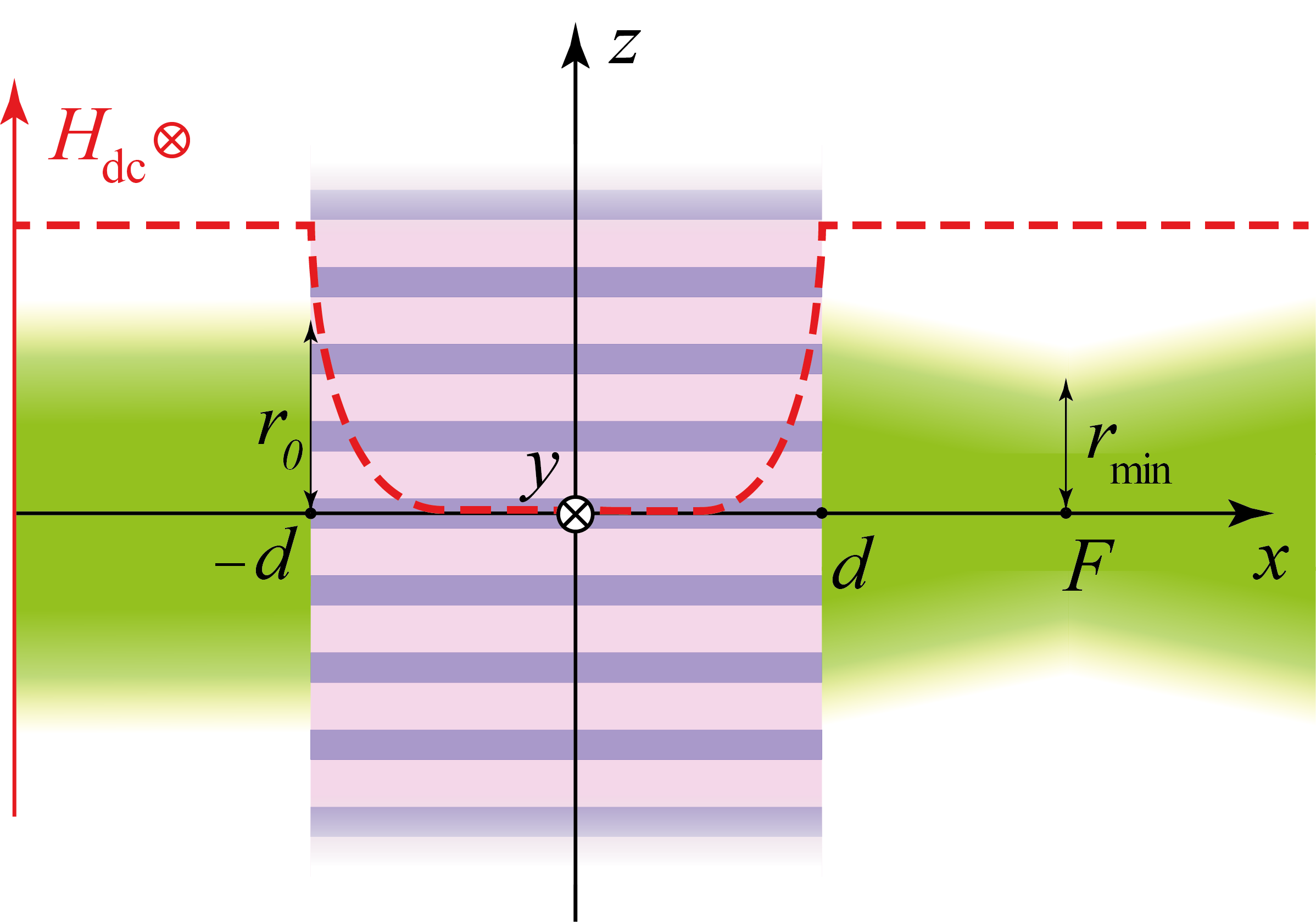}
\caption{\label{Figlas1}
{Schematics of the setup. A {parallel} laser beam of initial radius $r_0$ is focused by a plate of the layered superconductor, where $F$ and $r_{\text{min}}$ {correspond to} the focal distance and the waist radius of the transmitted beam, respectively.} Red dashed line shows schematically the spatial distribution of the DC field $H_{\rm dc}(x)$.
}
\end{center}
\end{figure}

In this paper, we study the transmission of a laser beam through a layered superconducting slab of thickness $2 d_{}$ in the external DC magnetic field {$\vec{H}_{\rm dc}=H_{\rm dc}\vec{e}_y$} in geometry presented in the Fig.~\ref{Figlas1}. {The superconducting layers of the slab are considered to be perpendicular to its interfaces revealing the anisotropic properties.} {The beam falls orthogonal to the surface of the superconducting sample and is polarised so} that the electric field is perpendicular to the layers, {$\vec{E}=E\vec{e}_z$}. In this polarization, only so-called extraordinary waves are induced in the slab which leads to {the maximally pronounced} nonlinear focusing of the transmitted beam~\cite{lasermain}. For simplicity reasons, we assume that the incident {beam is focused exactly} at the interface of the slab, resulting in zero curvature at the {left slab interface}, $x=-d_{}$.

We assume the laser beam to be Gaussian in vacuum and in the slab. Strictly speaking, due to multiple reflections as well as to nonlinearity, the laser beam should lose its Gaussian profile. However, for a sufficiently thin slab with realistic parameters, we {can
consider} the field inside the slab as a superposition of two Gaussian beams: ``forward'' and ``backward''. Acceptability of this assumption was verified by numerical simulation in~\cite{lasermain}. Below we briefly outline the main equations for the field in vacuum and in the layered superconductor.

\subsection{Gaussian Beam in Vacuum\label{sec:beam_vac}}

The distribution of the electric field for a Gaussian beam of frequency $\omega$ propagating along the $x$ axis in vacuum is as follows \cite{gaussvac1, gaussvac2}:
\begin{equation}
\label{evac}
 E(x,r,t)=E_v\exp\left(-\frac{r^2}{r_c^2}\right)\sin\Phi_v,
\end{equation}
with the phase {$\Phi_v$ depending on coordinates and time,}
\begin{equation}
\label{eqphigen}
\Phi_v(x,r,t)=k_v\left(x+\frac{\alpha r^2}{2}\right)-\omega t + \phi
\end{equation}
where $r^2=z^2+y^2$. The wave number $k_v$ obeys the dispersion relation for vacuum, $k_v=\omega/c$. The amplitude $E_0$, the radius $r_c$, the Gouy phase $\phi$, and the wavefront curvature $\alpha$ slowly change {along the beam path}, i.e. these characteristics of the beam are functions of $x$.

{Formally, Eqs.~\eqref{evac} and \eqref{eqphigen} describe the field in vacuum 
regions on both sides of the sample. For the region on the left (see Fig.~\ref{Figlas1}) we set $r_c=r_0$ and consider two beams, incident and reflected ones, with different amplitudes and phases. {In particular, $E_v=E_0$, $\alpha=\alpha_i=0$ and $\phi=\phi_i=0$
for the incident wave, while $E_v=E_r$, $\alpha=\alpha_r$ and $\phi=\phi_r$ are to be determined
for the reflected one. Subsequently,} for the transmitted wave the assumption of thin slab yields 
the condition $r_c=r_0$ on the right interface, the further decrease of $r_c$ corresponds to the focusing of the beam. Denoting the front curvature $\alpha=\alpha_t$ on the right interface, it} can be shown~\cite{gaussvac1} that {the transmitted beam converges to a minimum (beam waist) radius~$r_{\text{min}}$,
\begin{equation}
r_{\text{min}}=\frac{r_0}{\sqrt{1+\alpha_t^2 k_v^2 r_0^4/4}}
\label{eqrmin}
\end{equation}
at the focal distance $F$,
\begin{equation}
	F=-\frac{1}{\alpha_t+4/(\alpha_t k_v^2 r_0^4)},
	\label{eqF}
\end{equation}
see Fig.~\ref{Figlas1}.}

\subsection{Gaussian Beam in Layered
Superconductor \\ in the Presence of a Weak DC {Magnetic} Field}

\subsubsection{Electrodynamics of layered superconductor}

To describe the beam propagation inside the sample $(-d_{}<x<d_{})$ we employ the well-known coupled sine-Gordon equation for the phase difference of the order parameter $\varphi$ between neighboring layers in the continuum approximation, which can be presented as follows~\cite{Savel'ev_2010}:
\begin{equation}
	\label{SG}
	\left(1-\lambda_{ab}^2\frac{\partial^2}{\partial z^2}\right)\left[\frac{1}{\omega_J^2}\frac{\partial^2
		\varphi}{\partial t^2} +\sin\varphi \right] -
	\lambda_c^2\left(\frac{\partial^2\varphi}{\partial x^2}+\frac{\partial^2\varphi}{\partial y^2}\right)=0,
\end{equation}
where $\omega_J$ is the Josephson plasma frequency; $\lambda_{ab}$ and $\lambda_{c}=c/(\omega_J\sqrt{\varepsilon})$ are the London penetration depths along and across the layers, respectively; $\varepsilon$ is the dielectric constant of the interlayer gaps; $c$ is the speed of light. The electric and magnetic fields inside the sample can be expressed in terms of $\varphi$ \cite{lasermain}:
\begin{eqnarray}
	\label{eqEf}
E=-\frac{H_c\lambda_c}{2c}\frac{\partial \varphi}{\partial t},
\quad
\frac{\partial H}{\partial x}=-\frac{H_c}{2 \lambda_{c}}\left(\frac{1}{{\omega_J}^2} \frac{\partial^2 \varphi}{\partial t^2}+\sin \varphi\right).
\quad
\end{eqnarray}
Here $H_c=\Phi_0/(\pi s \lambda_{c})$ {is the critical field in layered superconductor},
$\Phi_0=\pi c \hbar/e$ is the magnetic flux quantum, and $s$ is the period of the layered structure.

\subsubsection{Nonlinear waves and DC field}

{As one can see, Eq.~\eqref{SG} is nonlinear in phase difference~$\varphi$. This means that the static magnetic field and the THz wave fields can not be independently added, which is of key importance in our problem. Instead,} we deal with the nonlinearity of Eq.~\eqref{SG} in the following way. We present~$\varphi$ in the form of superposition of a small term~$\varphi_w(x,r,t)$, which corresponds to the wave beam propagating in the layered superconductor {and \textit{depends on the static field}}, and the stationary solution~$\varphi_{s}(x)$ describing the DC magnetic field inside the slab~\cite{hmain}:
\begin{equation}
    \label{superpos}
	\varphi(x,r,t)=\varphi_{s}(x)+\varphi_w(x,r,t).
\end{equation}
{We consider static input in Eq.~\eqref{superpos} to be much greater than the wave one and} expand $\sin \varphi$, the single nonlinear term in  Eq.~\eqref{SG}, to the third order in small $\varphi_w$,
\begin{equation}
\label{eq:sin_phi_expand}
\sin\varphi= (1-\varphi_w^2/2)\sin\varphi_{s}+(\varphi_w-\varphi_w^3/6)\cos\varphi_{s}.
\end{equation}

The {main} term in Eq.~\eqref{eq:sin_phi_expand} (when neglecting small $\varphi_w^2$) with correspondent linear in $\varphi_{s}$ terms from Eq.~\eqref{SG} yields the following equation,
\begin{equation}
	\label{SG-stationary}
	\sin\varphi_s  -
	\lambda_c^2\frac{\partial^2\varphi_s}{\partial x^2}=0,
\end{equation}
which has a well-known solution describing the Josephson vortex. For relatively weak DC magnetic fields, $H_{\rm dc}<H_c$, and not too thin slabs, $\exp(-d_{}/\lambda_c)\ll1$,
we can expressed the stationary solution $\varphi_{s}$ as two ``tails'' of Josephson vortices,
\begin{eqnarray}
\label{eq:varphi_s}
	\varphi_{s}(x)&=&
 \varphi_0(x)-\varphi_0(-x),
 \\ \nonumber
\varphi_0(x)&=&4 \arctan[e^{-\xi_0-(d-x)/\lambda_{c}}],
\quad
\xi_0=\arccosh\frac{H_c}{H_{\rm dc}}.
\end{eqnarray}

The {terms containing $\varphi_w$ in Eq.~\eqref{eq:sin_phi_expand} describe the wave input in the solution. Generally, the $\varphi_w^3$ term corresponds to the \textit{weak} nonlinear effect, because $|\varphi_w|\ll 1$}. However, when frequency~$\omega$ is close enough to $\omega_J$,
\begin{equation}
	\label{beta}
\varphi_w \sim \beta=\sqrt{\omega^2/\omega_J^2-1}\ll1,
\end{equation}
nonlinear term $\varphi_w^3$ becomes crucial. {In Refs.~\cite{hmain, PhysRevB.82.144521, PhysRevB.78.184504, PhysRevB.90.184503}, a number of \textit{strong} nonlinear effects were predicted in this regime, when} {linear terms in Eq.~\eqref{SG} nearly cancel each other and the cubic term becomes of the same order as their difference}. In the present study, we imply exactly this frequency range to predict the nonlinear focusing of the beam. Substituting expansion~\eqref{eq:sin_phi_expand} into Eq.~\eqref{SG} and assuming condition~\eqref{beta}, we obtain the following differential equation for $\varphi_w$:
\begin{equation}
	\label{betaeq}
	\beta^2 \varphi_w+\frac{\varphi_w^3}{6}+(1-\cos \varphi_{s})\varphi_w+\lambda_{c}^2\frac{\partial^2 \varphi_w}{\partial x^2}=0,
\end{equation}
which is solved in the following section.

Note that in the latter equation, we neglect the second derivatives over $y$ and $z$. They turn out to be small in comparison to the other terms in Eq.~\eqref{SG}, when the
slab is thin compared to the radius of the beam, $d\ll r_0$, see details in Ref.~\cite{lasermain}.

\subsubsection{Gaussian beam}

{The beam propagating in the superconducting slab reflects on each of the two interfaces, which, strictly speaking, leads to the multiple reflections and the possible} loss of the Gaussian profile of the resulting beam.
However, as was shown in Ref.~\cite{lasermain} for a sufficiently thin sample, {$d\ll r_0$, the radius of the beam passing through the plate is preserved, and, therefore, all internally reflected beams propagating in one direction merge to a single beam. }

Under this assumption, 
we can represent
the electric field in the sample as a superposition of two Gaussian beams, propagating forwards (index “$+$”) and backwards (index “$-$”),
\begin{align}
\label{eqExrt}
E_S(x,r,t)&=H_c\exp\left(-r^2/r_0^2\right)\\&\times\left[E_+(x) \sin\Phi_ +(x,r,t)+E_-(x) \sin\Phi_-(x,r,t)\right],\nonumber
\end{align}
with phases
\begin{equation}
\label{eqphi}
\Phi_\pm(x,r,t)=\pm k_\pm(r)\Big[x+\frac{\alpha_\pm(x) r^2}{2}\Big]-\omega t + \phi_\pm(x).
\end{equation}

In the following section, we derive expressions for amplitudes $E_\pm$, curvatures $\alpha_\pm$, and phases $\phi_\pm$, as well as for the wave numbers $k_\pm(r)$.

\section{\label{sec:level3}EFFECT OF DC MAGNETIC FIELD ON LASER BEAM FOCUSING}

\subsection{Analytical approach}

Here we derive the analytic expression for the curvature~$\alpha_t$ of the transmitted beam as well as the relations between the amplitudes of the beam inside and outside the slab of layered superconductor. In the first step, we describe the beam deeply inside the slab where the DC magnetic field is negligible and we can exploit the approach developed in Ref.~\cite{lasermain} for the nonlinear beam propagation. Secondly, we extend that approach to the areas near the slab's surfaces and take into account the DC magnetic field. Finally, we match the tangential components of the wave electric and magnetic fields at the interfaces and find the result.

\subsubsection{{Beam propagation in the slab far from the interfaces}}

As stated in the previous section, the DC magnetic field fades exponentially { with distance} from the surfaces of the slab, {penetrating in the form of} the ``tails'' of Josephson vortices, see Eq.~\eqref{eq:varphi_s}.  The DC {input $\varphi_s$ is} negligible, if $|\varphi_s|\ll\beta$, or
\begin{equation}
\label{eq:deep_inside}
\exp[-\xi_0-(d-|x|)/\lambda_c]\ll \beta,
\end{equation}
that is fulfilled at a distance of several $\lambda_c$ from the surfaces.

Inside this region, the parameters of the laser beam {described by} Eqs.~\eqref{eqExrt} and \eqref{eqphi} can be found from the solution of Eq.~\eqref{betaeq} by the approach developed in Ref.~\cite{lasermain} for the problem without a DC magnetic field:
\begin{eqnarray}
\label{eq:kpm}
&& k_\pm(r)=k_s+\frac{\kappa_\pm}{\lambda_{c}}-\frac{\gamma_\pm}{\lambda_{c}}\frac{r^2}{r_0^2},
\quad
\label{eqgamma}
\gamma_\pm=\kappa_\pm\frac{2\beta+\kappa_\pm}{\beta+\kappa_\pm},
\\
\label{eqkappa}
&& 8\kappa_\pm(2\beta+\kappa_\pm)=\varepsilon(E_\pm^2+2 E_\mp^2),
\end{eqnarray}
where $k_s$ obeys the dispersion law for linear waves in a superconductor:
\begin{equation}
{k_s^2\lambda_c^{2}=\omega^2/\omega_J^2-1.}
\end{equation}

Note that Eqs.~\eqref{eqgamma} and \eqref{eqkappa} imply that the amplitudes $E_\pm$, curvatures $\alpha_\pm$ and phases $\phi_\pm$ of the laser beam are constant inside the sample where the DC magnetic field is negligible. However, near the plate's surfaces, they can gain $x$-dependence, see Eq.~\eqref{Edep} and explanations nearby. {This means that despite the smallness of the penetration depth for the DC magnetic field, it still can be an effective instrument to tune the focusing of the layered superconductor slab.} The radial dependence of the wave number $k_\pm(r)$ in Eq.~\eqref{eq:kpm} arises from nonlinearity and means that different regions of the wavefront have different wave speeds, which leads to the fact that the beam can converge after passing through the plate.

\subsubsection{{Beam propagation in the vicinity of the slab interfaces}}

Near the boundaries of the sample, where the DC magnetic field penetrates as ``tails'' of the Josephson vortex, the factor $(1-\cos\varphi_{s})$ in Eq.~\eqref{betaeq} is not small, and therefore one can neglect first two terms in Eq.~\eqref{betaeq} due to the {smallness of} parameter~$\beta$. Substituting the electromagnetic field of the beam, Eq.~\eqref{eqExrt},  {to Eqs.~\eqref{eqEf} and~\eqref{betaeq}}, we get an equation that relates the beam's characteristics.

To solve that equation we expand it with respect {to} $r^2\ll r_0^2$ and keep only up to quadratic terms. The higher terms give contributions to the deviations from the Gaussian profile of the laser beam which are small in the nonlinear regime {defined by Eq.~\eqref{beta}}. Considering separately {terms with} $\cos\omega t$, $\sin\omega t$, $r^2\cos\omega t$, and $r^2\sin\omega t$, we get equations for the beam's characteristics. Solving these equations, we find:
\begin{eqnarray}
\label{Edep}
E_\pm(x)&=&f(x) E_\pm,
\,\,
\phi_\pm(x)=\phi_\pm,
\,\,
\alpha_\pm(x)=\alpha_\pm,
\\\nonumber
f(x)&=&f_0(x-d_{})+f_0(-x-d_{})+1,
\\\nonumber
f_0(x)&=&\tanh(x/\lambda_{c}+\xi_0)-1,
\end{eqnarray}
where $E_\pm$, $\phi_\pm$, and $\alpha_\pm$ are constant {far from the sample interfaces}, where the DC magnetic field is small.

Note that Eqs.~\eqref{Edep} are given in the main approximation with respect to the small parameter~$\beta$ and for the not very thin sample, $d\gg\lambda_c$~\cite{hnonlinearmain}.
Indeed, e.g. for the phase, we can get more accurate results than in Eq.~\eqref{Edep}:
{
\begin{equation}
\phi_\pm'(x)=\frac{k_{\pm0}}{f^{2}(x)}-k_{\pm0},
\quad
k_{\pm0}=k_{\pm}(r=0),
\label{eq:phi_pm}
\end{equation}
}
where the dash denotes the derivative with respect to the variable $x$.
Although the phase itself depends on $x$, the phase deviations $\phi_\pm(d)-\phi_\pm(-d)$ over the entire sample is of order of small parameter~$\beta$ and can be neglected:
{
\begin{equation}
\int_{-d_{}}^{d_{}} \phi_\pm'(x) dx = \int_{-d_{}}^{d_{}} \left(\frac{k_{\pm0}}{f^{2}(x)}-k_{\pm0}\right) dx  \sim \lambda_{c}k_s = \beta \ll 1.
\end{equation}
}
Similarly, the same can be shown for the curvatures $\alpha_\pm(x)$. So, we can consider the phases $\phi_\pm$ and curvatures $\alpha_\pm$ to be constant along the $x$ coordinate as stated in Eq.~\eqref{Edep}.

For the magnetic component of the beam field (directed along the $y$ axis), we derive from Eq.~\eqref{eqExrt} using Eqs.~\eqref{eqEf} and~\eqref{Edep} the following expression,
\begin{equation}
\label{eqHxrt0}
H_S(x,r,t)=H_{S+}(x,r,t)+ H_{S-}(x,r,t),
\end{equation}
where
{
\begin{eqnarray}
\label{eqHxrt}
&&H_{S\pm}(x,r,t)=
H_c H_\pm\exp\left(-r^2/r_0^2\right)
\\&&\notag\quad\times
\Big[
\dfrac{k_\pm(r)}{k_{\pm0}f(x)}
\sin\Phi_\pm(x,r,t)\pm
\dfrac{f'(x)}{k_{\pm0}}
\cos\Phi_\pm(x,r,t)\Big],
\\ \nonumber
&&\qquad
H_\pm=\pm \sqrt{\varepsilon}\lambda_{c} k_{\pm0} E_\pm.
\end{eqnarray}
}
Note that the first and second summands in the square brackets are calculated using Eqs.~\eqref{eq:phi_pm} and~\eqref{Edep}, respectively.

\subsubsection{Matching {of the fields} at the interfaces}

Now we relate laser beams {in vacuum regions, incident and reflected ones for $x=-d$, and a transmitted one for $x=d$}, with the beams inside the superconducting plate. For this purpose, we match the tangential components of the electric and magnetic fields at the two interfaces between the vacuum and the superconductor.

The quantities sought in the problem are amplitudes $E_{r/t/\pm}$, phases $\phi_{r/t/\pm}$, and curvatures $\alpha_{r/t/\pm}$ of the beams propagating {in forward and backward directions both in vacuum and superconductor regions, 12~quantities in total}. The parameters of the incident beam are assumed to be known, $E_{i}$, $\phi_{i}=0$, and $\alpha_{i}=0$. To obtain a set of 12~equations for the required quantities, we expand the expressions for the fields into series in radius $r$ and keep the terms up to the second order, because we neglect the non-Gaussian deviations produced by the higher order terms. Moreover, since $k_\pm\lambda_c\sim \beta$ and $\sqrt{\varepsilon}\sim 1$, the amplitudes of the magnetic field are much smaller than those of the electric field,
\begin{equation}
\label{hlle}
H_\pm \sim \beta E_\pm \ll E_\pm.
\end{equation}
Therefore, we neglect the terms of order of $r^2$ in magnetic field and keep only the main order.

As a result, we get the required set of 12~equations. Eliminating the phases~$\phi_{r/t/\pm}$, curvatures~$\alpha_{r/\pm}$, and amplitudes~$E_{r/t}$, we derive two equations relating amplitudes~$E_\pm$ with the incident amplitude~$E_i$,
\begin{eqnarray}
\label{simple}
    &&\chi E_++H_+=-(\chi E_-+H_-),\\ \nonumber
    &&\sqrt{\chi}\frac{E_+-E_-}{2}\sin(2\bar{\kappa} d_{}/\lambda_c)=\frac{E_i}{H_c},
\end{eqnarray}
and expression for the curvature~$\alpha_t$ of the transmitted beam,
\begin{equation}
\label{alphateq}
\alpha_t=\frac{2 d_{}}{k_v \lambda_{c}r_0^2 }
\Big\{\tilde{\gamma}-\frac{\bar{\gamma}\left[(1-h_0^2)/\sqrt{\varepsilon}+\tilde{\kappa}\right]}{\bar{\kappa}-h_0^2 \sqrt{1-h_0^2}\tan (2\bar{\kappa}d_{}/\lambda_c)}\Big\},
\end{equation}
where $\bar{\gamma}=(\gamma_++\gamma_-)/2$,
$\tilde{\gamma}=(\gamma_+-\gamma_-)/2$,
\begin{equation}
    \bar{\kappa}=\dfrac{\kappa_++\kappa_-}{2}+\beta,
    \quad
    \tilde{\kappa}=\dfrac{\kappa_+-\kappa_-}{2},
\end{equation}
and factor $\chi$ depends only on the DC field magnitude:
\begin{equation}
\chi=1-h_0^2+\varepsilon h_0^4,
\quad h_0=H_{\rm dc}/H_{c}.
\end{equation}

The parameters~$\kappa_\pm$, magnetic~$H_\pm$ and electric~$E_\pm$ field amplitudes are related by a set of nonlinear algebraic equations~\eqref{eqkappa}, \eqref{eqHxrt}, and~\eqref{simple}. Solving this set {of equations} we can establish value of curvature~$\alpha_t$, Eq.~\eqref{alphateq}, and, subsequently, the beam waist radius~$r_{\rm min}$ and focusing length~$F$, Eqs.~\eqref{eqrmin} and~\eqref{eqF}, as nonlinear functions of the amplitude~$E_i$ of an incident beam and the DC field magnitude~$H_{\rm dc}$ (via $h_0$).

\subsection{The results of analytic approach}

{Thus, the developed theoretical approach allows to obtain} the focal distance $F$ and the beam waist radius $r_{\text{min}}$ for the transmitted beam {from Eqs.~\eqref{eqrmin}, \eqref{eqF} and~\eqref{alphateq}. Now we proceed with the analysis of how the quantities $F$ and $r_{\text{min}}$ depend on key parameters of the problem, frequency $\omega$, amplitude $E_0$ of incident beam and DC field magnitude $H_{\rm dc}$.}

\begin{figure}[h]
	\begin{center}
		\includegraphics[width= 8.5 cm]{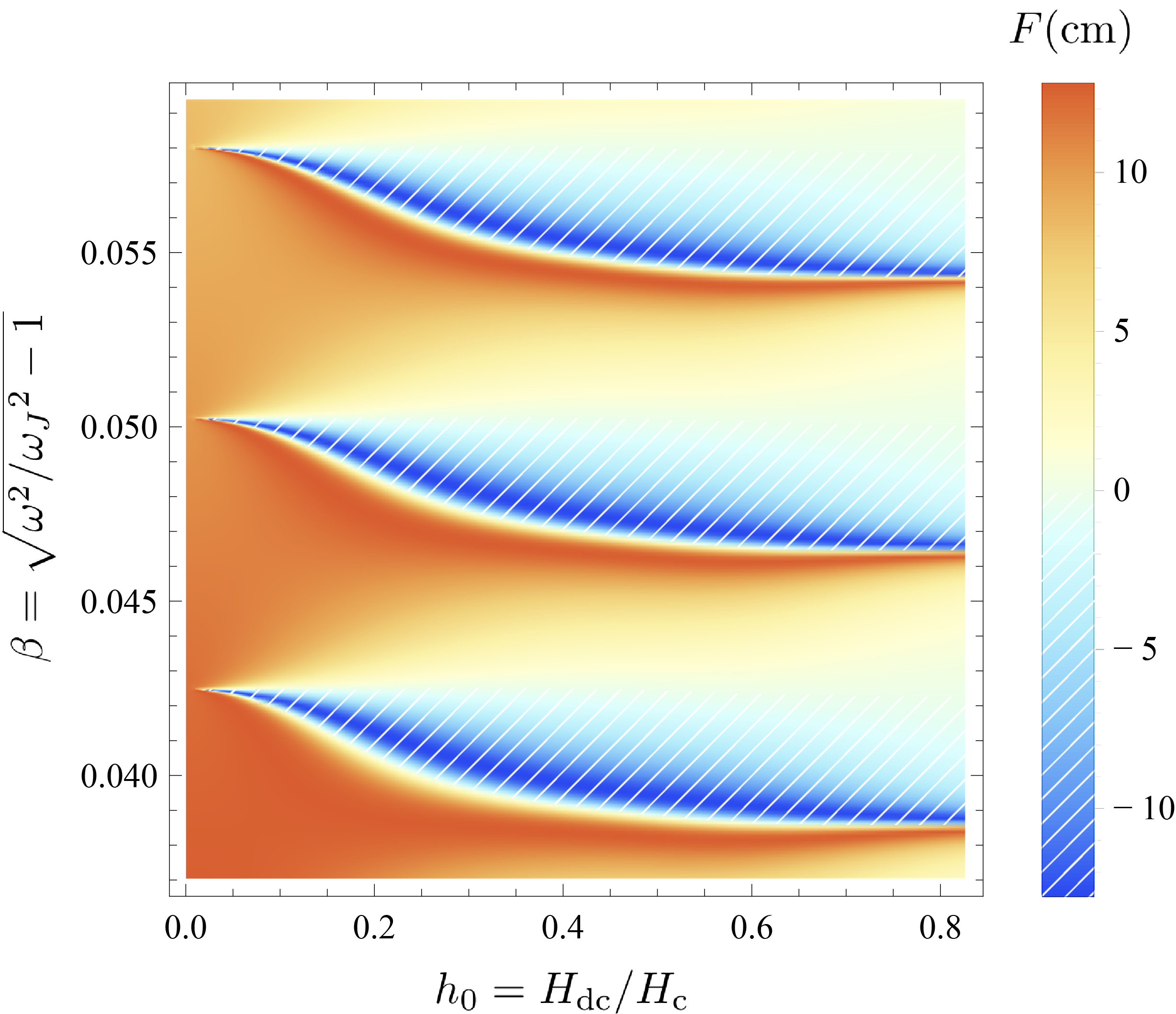}
		\caption{\label{figlasF} (Color online) The dependence of the focal length $F$ on the dimensionless frequency represented by parameter $\beta=\sqrt{\omega^2/{\omega_J}^2-1}$ and the magnetic field magnitude $h_0~=~H_{\text{dc}}/H_c$ as obtained from Eqs.~\eqref{eqrmin}, \eqref{eqF} and~\eqref{alphateq}. Shaded areas correspond to defocusing. {The strong dependence on the DC magnetic field allows us to use it to control the focusing of THz radions.} Parameters are: $\omega_J/2\pi$~=~2~THz, $\gamma$~=~15, $r_0$~=~3.5~mm, $2d_{}~=~2.5$~mm, s~=~20~\r{A}, $E_i~=~0.05$~kV/cm, $\varepsilon$~=~15. }
	\end{center}
\end{figure}

Fig. \ref{figlasF} shows the dependence of the focal length $F$ on the frequency detuning from the Josephson plasma frequency, represented by the dimensionless parameter $\beta$, and on the parameter $h_0$, which represents the normalized DC magnetic field. It can be seen that {despite the fact that external DC magnetic field penetrates the slab only by the small depth $\sim\lambda_c$, it} significantly affects the focal length. {This means that it can be used to tune the focusing of THz radiation. As for the frequency dependence, we see that for a non-zero field, there is an almost periodic variation with the change of} parameter $\beta$. The ``period'' in $\beta$ can be estimated from simple considerations. The periodicity is determined by the number of wavelengths that fit in the sample, which means that the periodicity in $k_x$ satisfies $2 d_c \Delta k_x = 2 \pi$. Since $k_x\approx\beta/\lambda_{c}$, we get
\begin{equation}
\Delta \beta \approx \frac{\pi \lambda_{c}}{d_c}.
\end{equation}
This estimation for the parameters used in Fig.~\ref{figlasF}, gives $\Delta \beta\approx 0.008$, which corresponds well to what the analytic calculation gives.

{It is interesting to note that the shaded blue regions in Fig.~\ref{figlasF} correspond to the negative focal distance, which means \textit{defocusing} of the beam. Meanwhile, in the absence of external magnetic field the superconducting slab can \textit{only focus} radiation due to nonlinearity~\cite{lasermain}. This broadens the variety of control possibilities in the system.}

\begin{figure}[h]
	\begin{center}
		\includegraphics[width= 8.5 cm]{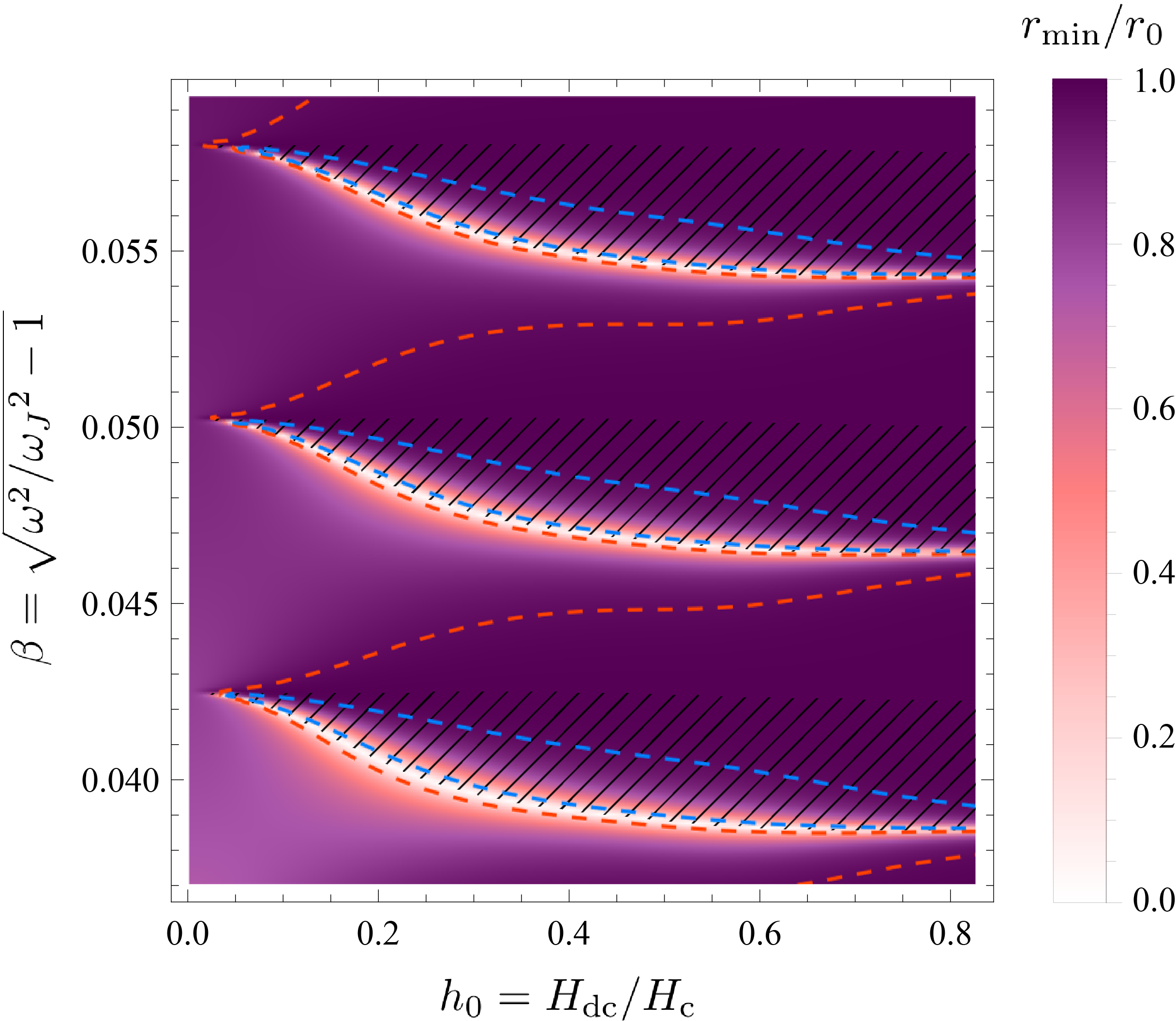}
		\caption{\label{figlasr2} (Color online) Dependence of the normalized beam waist radius $r_{\text{min}}/r_0$ on frequency detuning $\beta=\sqrt{\omega^2/{\omega_J}^2-1}$ and DC magnetic field $h_0~=~H_{\text{dc}}/H_c$. Shaded areas correspond to defocusing. Parameters are the same as in Fig. \ref{figlasF}.}
	\end{center}
\end{figure}

{Consider now the beam waist radius $r_{\text{min}}$ as the function of frequency and DC magnetic field magnitude. Figure~\ref{figlasr2} shows the dependence on these parameters. The red and blue dashed lines correspond to the constant focal lengths of 6 cm and $-6$ cm, respectively.} Note the sharp increase in focusing efficiency near the regions where the focal distance changes its sign.

\begin{figure}[h]
	\begin{center}
		\includegraphics[width= 8.5 cm]{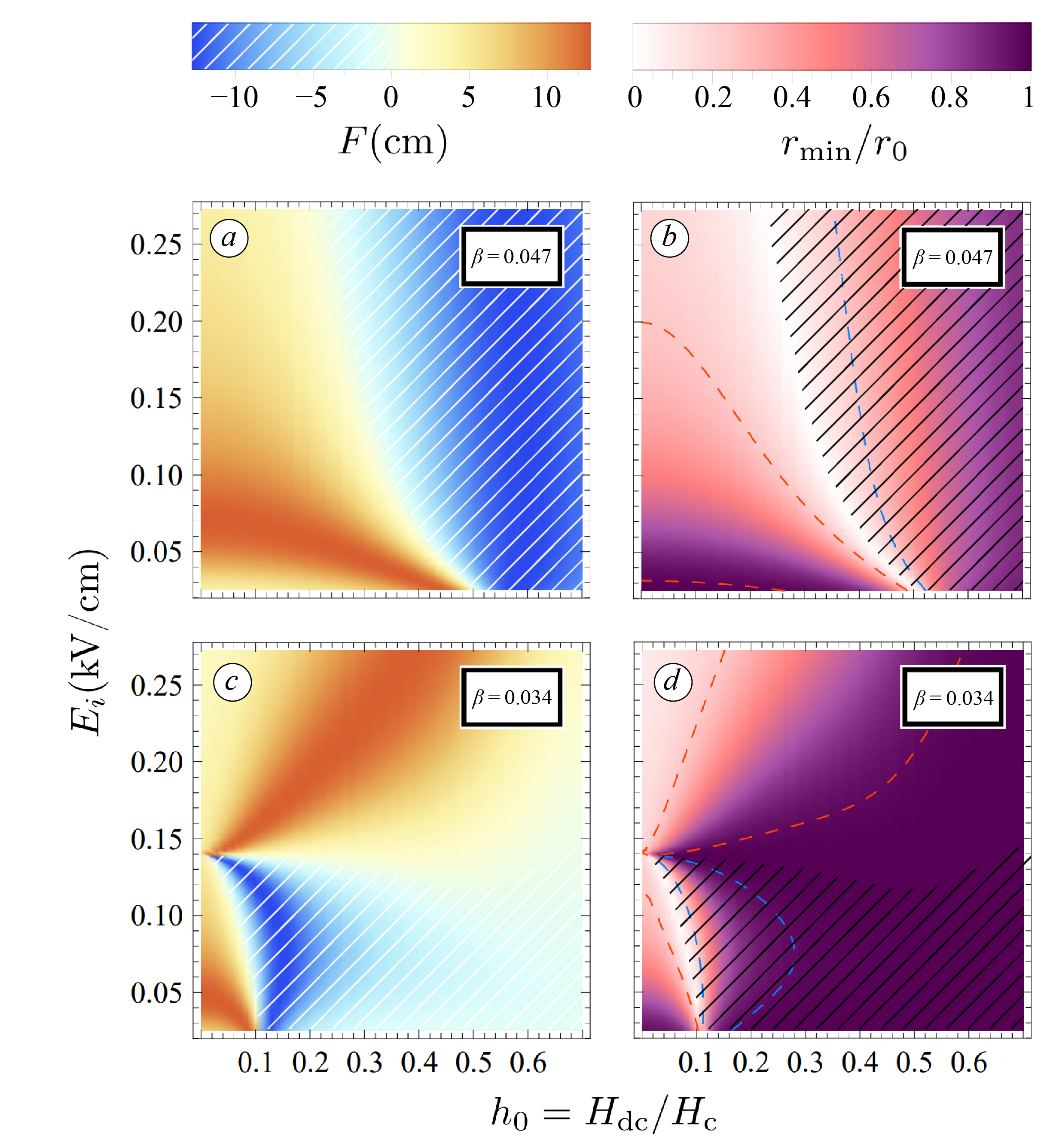}
		\caption{\label{figlas41} (Color online) Dependence of the focal length $F$ and normalized beam waist radius $r_{\text{min}}/r_0$ on the amplitude $E_i$ of the incident wave and DC magnetic field $h_0~=~H_{\text{dc}}/H_c$. Shaded areas correspond to defocusing. Parameters are the same as in Fig. \ref{figlasF}.}
	\end{center}
\end{figure}

{Finally, as was discussed, the focusing ability of the layered superconductor relies on non-linear equations for the wave field. This obviously implies that the focusing parameters depend on the amplitude of the incident beam. We demonstrate this in~Fig. \ref{figlas41}, where we show the dependences of focal distance $F$ and the beam waist radius $r_{\text{min}}$ on the amplitude $E_i$ of the incident beam and the magnitude of the external DC magnetic field. We see that tuning the focusing parameters by both beam amplitude and external magnetic field requires much less precision than tuning by frequency, and thus may be profitable in applications. Moreover, if the task is to focus the THz beam of a given frequency and amplitude, the external magnetic field becomes a handy way to achieve this.}

\subsection{Numerical simulation\label{sec:numsim}}

{To verify the presented analytic results, we conducted a numerical simulation of the beam field in the layered superconductor slab.} The motivation behind was to check the assumptions that (i) the beam preserves a Gaussian profile in both the $y$ and $z$ directions; (ii) the higher-order terms in sine-Gordon equation and higher harmonics are negligible. So the comparison of the results obtained from simulation and analytic theory determines whether these effects potentially influence the entire beam propagation.

The simulation consisted in solving the sine-Gordon equation~\eqref{SG} using the Euler method for the phase difference $\varphi$ as a function of the coordinates $x,~y,~z$ inside the layered superconductor slab, as well as time $t$. The sizes of sample in the $y$ and $z$ directions were chosen to be $N=3$ times larger than the diameter of the beam. These sizes were sufficient to neglect the edge effects, as the electromagnetic field on the edge was $\exp(N^2)\sim 10^4$ weaker than on the axis of the beam. The mesh inside the slab was rectangular with a constant step size. In each direction, we chose an optimal step size, defined {from the condition that the sum of rounding and approximation errors was minimal.}

\begin{figure}
\includegraphics[width= 8.3 cm]{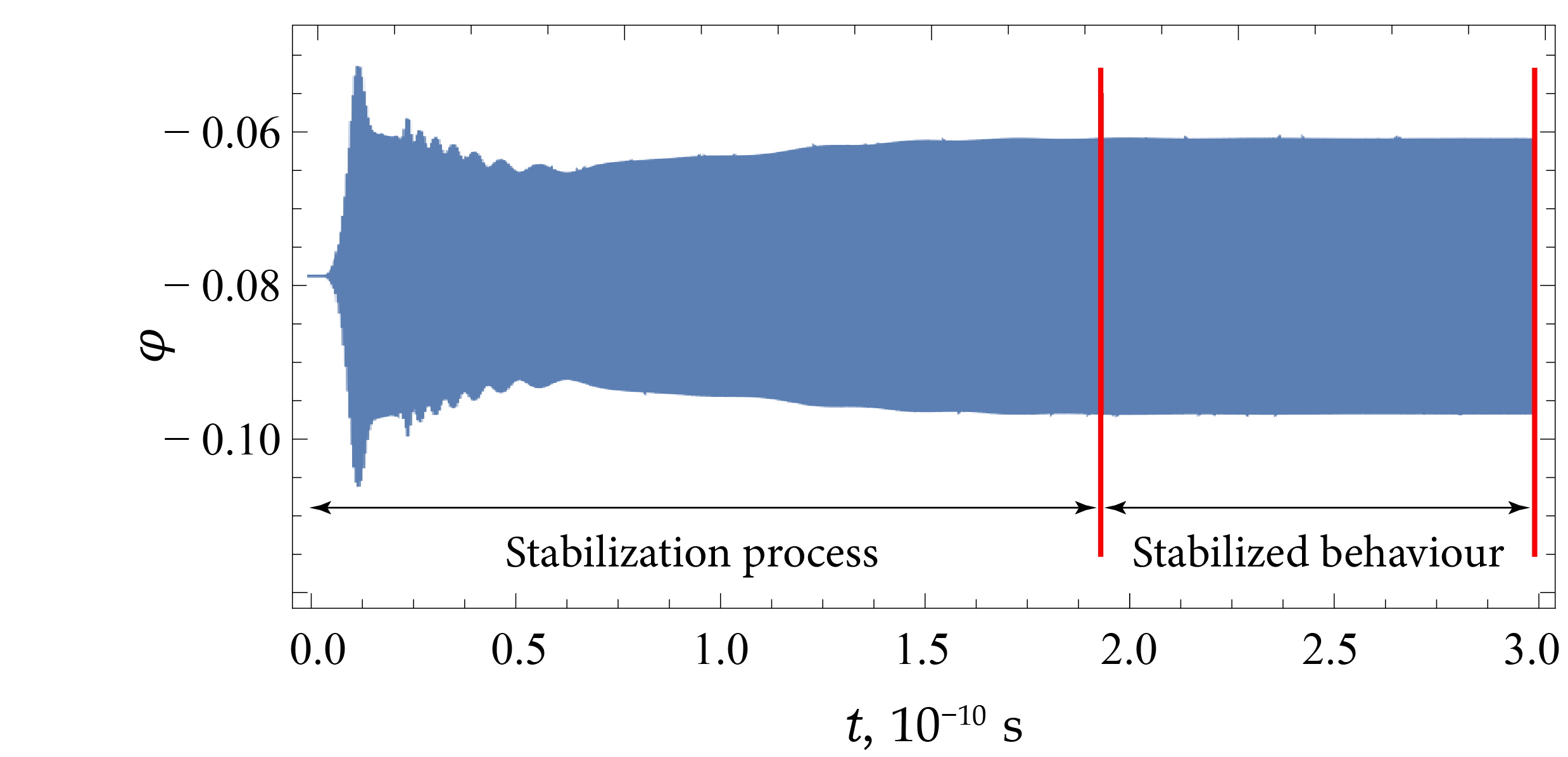}
\caption{\label{fig:simul_ampl_stab} Typical dependence of phase difference $\varphi$ at the beam axis on the right interface of the layered superconductor slab on time obtained from the simulation. This plot is schematically divided into a stage of stabilization process and a stage of stabilized behavior. Parameters for this plot: {$h_0=H_{\rm dc}/H_0=0.04$, normalized amplitude $E_0/H_c=0.05$, frequency detuning $\beta=1.1*10^{-3}$, other parameters are the same as in Fig. \ref{figlasF}}}
\end{figure}

Before simulating the beam propagation, we established the stationary phase $\varphi_{s}$ inside the slab according to Eq.~\eqref{eq:varphi_s}, which was determined by DC magnetic field. After that, we solved the sine-Gordon equation~\eqref{SG} with the corresponding boundary conditions for the field on two interfaces. The incident beam amplitude was constant, while the field values for reflected and transmitted beams on the interfaces were computed on each step of simulation until the amplitude of oscillations of phase $\varphi$ in each point established. The typical time evolution is shown in Fig.~\ref{fig:simul_ampl_stab}, where one can see the stabilization stage. Note that physically, this stabilization corresponds to several reflections between the left and right interfaces before the boundary conditions are met. After the stabilization was verified by demanding that the variation of oscillations amplitude was less than the chosen threshold value, we performed Fourier transform for the transmitted wave at each point on the right interface. We verified that the higher harmonics are indeed negligible (the typical fraction of power in them did not exceed $2\%$ for our calculations), and the peak value for frequency of incident wave allowed to extract the phase of the transmitted radiation. By fitting the distribution of phase at different points on the right interface with parabolic profile, we obtained the curvature of the transmitted beam, which was compared with the results obtained from our analytical approach. We also fitted the field amplitude distribution in the beam cross section with Gaussian and estimated the deviation from it to be less than $3\%$ for most of our calculations (see more details in Ref.~\cite{lasermain}).

\begin{figure}[h]
	\begin{center}
		\includegraphics[width= 8.5 cm]{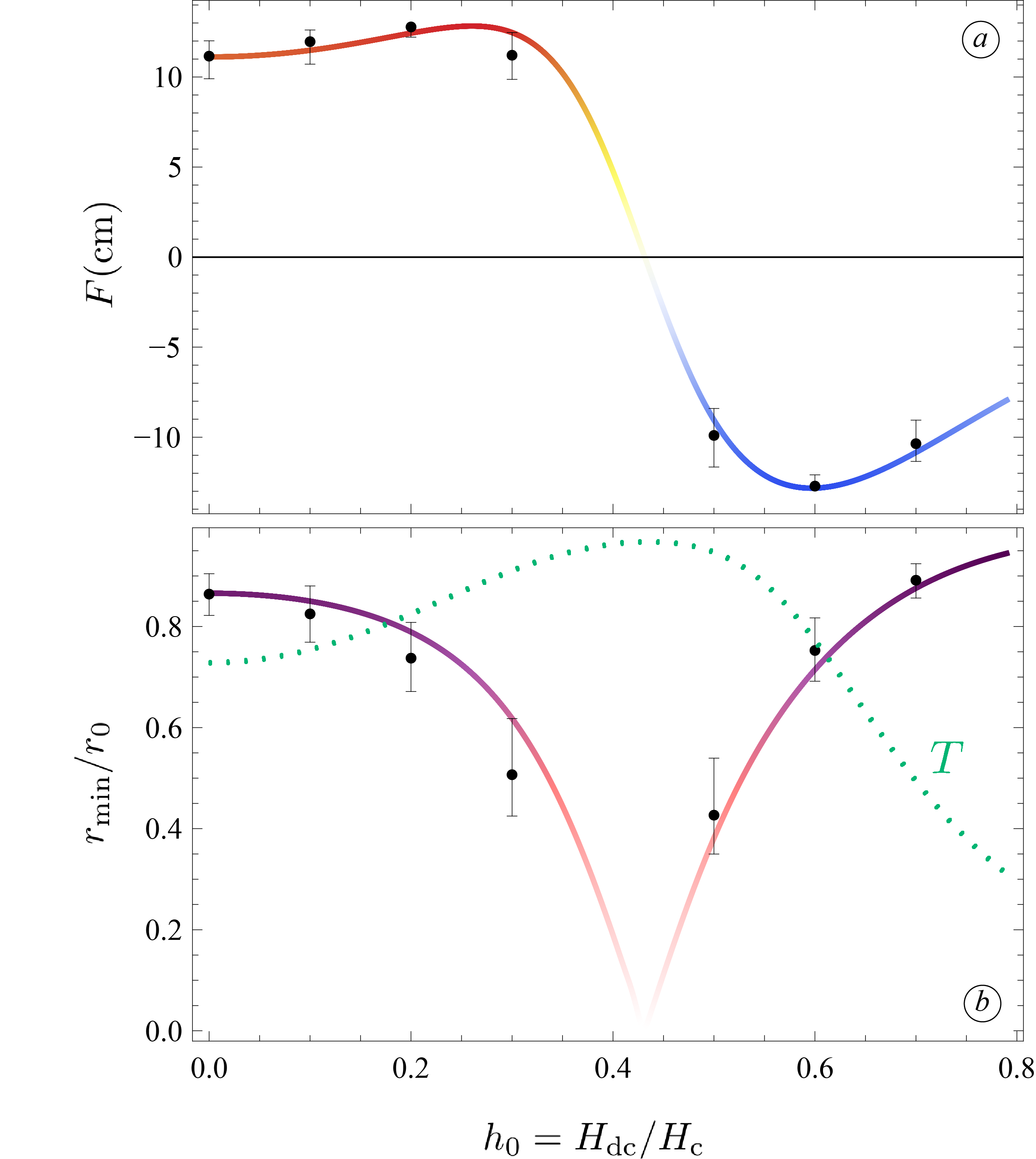}
		\caption{\label{figlassim} (Color online) Comparison of curves obtained analytically (solid lines) and results of numerical simulation (black dots) for the dependences of focal length $F$ (a) and normalized beam waist radius $r_{\text{min}}/r_0$ (b) on the magnetic field $h_0~=~H_{\text{dc}}/H_c$. {Color of the lines corresponds to the color schemes in Figs.~\ref{figlasF} and \ref{figlasr2}.} Green dotted line is the analytical curve for transmission coefficient $T$ (the scale is the same as for the normalized beam width). Parameters are the same as in Fig. \ref{figlasF}, $\beta=0.047$.}
	\end{center}
\end{figure}

{Figure~\ref{figlassim} shows the dependences of focal distance and beam waist radius on the external magnetic field and demonstrates the comparison of the simulation and analytic theory. As one can see, the results match each other pretty well for the most of the points.} For magnetic fields where the focal distance is too small, we see the deviations due to several reasons: (i) computation error strongly increases because in this case the characteristic lengths become comparable with the step size, (ii) beam {converges inside the slab and changes its shape, that does not allow to properly find the curvature of the transmitted wave. These reasons violate the conditions of applicability of our analytic results. However, one can see that the critical DC magnetic field for which the strong focusing takes place is still qualitatively predicted by the analytic theory.}

{Green dotted line in Fig.~\ref{figlassim} shows the results of our calculation of beam transmittance through the slab, which we could compute using our theoretical results. One can see, that in the region where the focusing of the slab is strong, the transmission is rather high. This means, that the THz beam focusing by the layered superconductor slab is indeed effective for different applications, compare to transmission in the range from 0.5 to 0.8 for Teflon and silicon lens~\cite{Mendis2016}.} Note, that the dissipative quasiparticle current in the superconducting slab leads to the decay of the beam as it propagates and multiply reflects from the interfaces. For the realistic parameters reachable in experiments~\cite{PhysRevB.68.134504,PhysRevB.64.174508}, the dissipation relaxation rate $\omega_r$ may be as low as $(10^{-4}\div 10^{-3})\omega_J$. Then the attenuation length can be estimated to be as high as several centimeters for frequency detuning ${\omega-\omega_J=10^{-3}\omega_J}$, which allows us to omit the dissipation for the samples with thicknesses about millimeters. Note also that the relaxation rate is highly sensitive to the temperature and may be reduced significantly by cooling the system.

\section{CONCLUSIONS}

We have developed theory of the propagation of a Gaussian laser beam through the layered superconductor plate placed in an external DC magnetic field. {Due to the specific non-linearity of equations describing electromagnetic field in the slab, it can focus Gaussian beam.} The DC magnetic field penetrates the plate non-uniformly and creates a non-linear background for the propagation of the THz beam in the plate that allows tuning the focusing parameters of the plate by varying the external DC field.

We show that depending on the parameters of the problem, a non-zero magnetic field can lead to both strong focusing and defocusing of the laser beam. The use of the DC magnetic field can be a more
convenient way of tuning the focusing of the laser beam than by changing its frequency. Presented analytic calculations are confirmed by numerical simulation.

\section{ACKNOWLEDGMENTS}
We gratefully acknowledge support from
the National Research Foundation of Ukraine, {Project No. 2020.02/0149} “Quantum phenomena in
the interaction of electromagnetic waves with solid-state
nanostructures”.

\bibliography{apssamp}

\end{document}